\documentclass{JINST}
\usepackage{epsfig}

\title{Monolithic Pixel Sensors in Deep-Submicron SOI Technology}

\author{Marco Battaglia$^{a,b}$, Dario Bisello$^c$,
Devis Contarato$^b$\thanks{Corresponding author.}, Peter Denes$^b$, Piero Giubilato$^{b,c}$,
Lindsay Glesener$^a$, Serena Mattiazzo$^c$, Chinh Qu Vu$^b$\\
\llap{$^a$}University of California at Berkeley,\\
  Berkeley, CA 94720 (USA)\\
\llap{$^b$}Lawrence Berkeley National Laboratory,\\
  Berkeley, CA 94720 (USA)\\
\llap{$^c$}Istituto Nazionale di Fisica Nucleare, Sezione di Padova,\\
  Padova, I-35131, Italy \\
  E-mail: \email{DContarato@lbl.gov}}

\abstract{Monolithic pixel sensors for charged particle detection and imaging applications
have been designed and fabricated using commercially available, deep-submicron Silicon-On-Insulator
(SOI) processes, which insulate a thin layer of integrated full CMOS electronics from a high-resistivity
substrate by means of a buried oxide. The substrate is contacted from the electronics layer through vias
etched in the buried oxide, allowing pixel implanting and reverse biasing. This paper summarizes the
performances achieved with a first prototype manufactured in the OKI 0.15~$\mu$m FD-SOI process,
featuring analog and digital pixels on a 10~$\mu$m pitch. The design and preliminary results
on the analog section of a second prototype manufactured in the OKI 0.20~$\mu$m FD-SOI process
are briefly discussed.}


\begin{document}
\section{Introduction}
Silicon on insulator (SOI) technology allows the integration of CMOS electronics
on a thin silicon layer which is electrically insulated from the wafer substrate
by means of a buried-oxide (BOX). With respect to bulk CMOS processes, the small
active volume ensures latch-up immunity and the low junction capacitance enables
low-threshold and low-noise operation, thus favoring high-speed and low power
digital designs.

SOI Monolithic active pixel sensors for charged particle detection and imaging
applications may be built by providing a technology to contact a high-resistivity
substrate through the BOX, thus allowing the formation of pixel and biasing implants.
With respect to conventional CMOS pixel sensors fabricated on undepleted epitaxial
layers, a full CMOS circuitry can be integrated in each pixel and the substrate
can be reverse biased and thus depleted for improved charge collection efficiency.
A proof of principle of the concept was demonstrated by the SUCIMA Collaboration using
a non-commercial, 3~$\mu$m feature size process from IET, Poland~\cite{marczewski2005,marczewski2006,niemec2006}.
Commercial deep-submicron, fully-depleted (FD) SOI CMOS processes were recently made
available by OKI, Japan~\cite{oki} through a collaboration with KEK, allowing for the
integration of complex architectures in devices with small pixel pitch. The CMOS electronics
is implanted on a 40~nm thin Si layer, which is fully depleted at typical operational
voltages, on top of a 200~nm thick BOX contacting a 700~$\Omega\cdot$cm Si substrate.
A 0.15~$\mu$m process was first validated for the fabrication of SOI pixel sensors by
KEK in 2006~\cite{tsuboyama2007}. Two prototype pixel sensors, featuring both analog
and digital pixels have been subsequently designed at LBNL and fabricated in OKI
processes. The experimental test results and the performances achieved with the LDRD-SOI-1
chip~\cite{battaglia_nima}, fabricated in 2007 in the 0.15~$\mu$m process, will be summarized
in Section~\ref{soi1}. Tests of the analog and digital pixels include in-lab studies
of charge collection with infrared lasers and high-energy particle beam tests on a 1.35~GeV
electron beam at the LBNL Advanced Light Source. The effects of ionising and non-ionising
radiation doses on the chip operation have also been studied. The LDRD-SOI-2 chip was then
fabricated in 2008 using a 0.20~$\mu$m process, featuring a larger pixel array and a more
complex architecture, optimized for faster readout. The chip design and first results
obtained on the analog pixels will be presented in Section~\ref{soi2}.

\section{Prototype in 0.15~$\mu$m Process}\label{soi1}
\subsection{Chip Design and Experimental Setup}\label{soi1_design}
The LDRD-SOI-1 chip features an array of 160$\times$150 pixels on a 10~$\mu$m pitch, subdivided
into two analog sections, with thin oxide 1.0~V transistors and thick oxide 1.8~V transistors,
and one digital section with 1.8~V bias. The analog pixels implement a simple 3-transistor (3T)
architecture, while each digital pixel implement a latch triggered by a clocked comparator, driven
by a dedicated readout clock and referenced to an adjustable voltage threshold which is common to the
whole matrix. In order to avoid static power dissipation, no amplifier is present in the digital
pixels. Like CMOS logic, the comparator consumes power only on clock transitions.

The reverse bias of the sensor substrate increases the potential at the silicon surface, and
causes a back-gating effect as the buried oxide may act as a second gate for the CMOS electronics
on top, typically causing a shift in the transistor thresholds with increasing depletion voltage.
The test of single $p$-type and $n$-type MOSFETs integrated at the chip periphery showed a
threshold shift of $\sim$200~mV for a substrate bias of 15~V. The effect had been investigated during
the design process by means of TCAD simulations, performed with the Synopsys {\tt Taurus Device}
package. The electrostatic potential at the interface between the buried oxide and the silicon substrate
was simulated as a function of the substrate bias for different pixel layouts, varying the size of the
charge collecting diodes and evaluating the effect of different guard-ring configurations. Simulation
showed that the inclusion of a floating $p$-type guard-ring around each pixel is beneficial in keeping
the field low in the area between diode implants, thus limiting possible back-gating effects on the
CMOS electronics on top of the buried oxide. A series of floating and grounded guard-rings were also
implemented around the pixel matrix and around the peripheral I/O electronics.

The chips sections are read out independently by a custom FPGA-based readout board with 14-bit ADCs,
whose digital outputs are connected to a National Instruments digital I/O PCI board installed on a
control PC running a LabView-based data acquisition and on-line data processing program. Pixels are
read out at 6.25~MHz clock frequency, with an integration time of 1.382~ms for the analog pixels.
Correlated Double Sampling (CDS) is obtained from the subtraction of two frames acquired without resetting
the pixels in-between. The integration time for the digital pixels is tunable, and their binary
output is buffered through the FPGA to the DIO board on the control PC. Data is stored in the {\tt lcio}
format~\cite{Gaede:2003ip}. Offline data analysis is based on a set of dedicated processors developed in
the {\tt Marlin} C++ reconstruction framework~\cite{Gaede:2006pj}.

\subsection{Analog Pixels}
\begin{figure}[t]
\begin{center}
\begin{minipage}{.49\linewidth}
\centerline{\epsfig{file=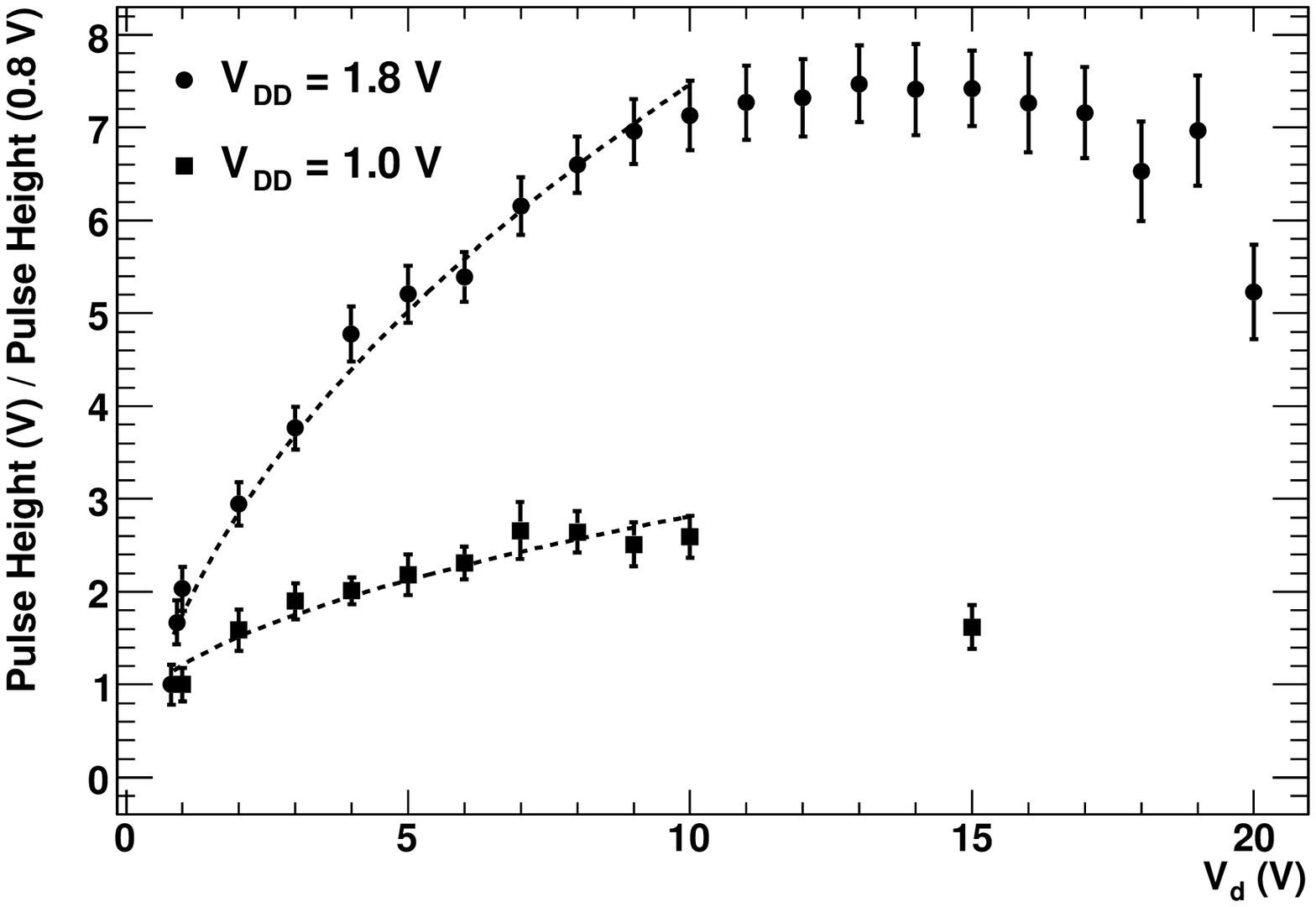,width=7.0cm}}
\end{minipage}
\hfill
\begin{minipage}{.49\linewidth}
\centerline{\epsfig{file=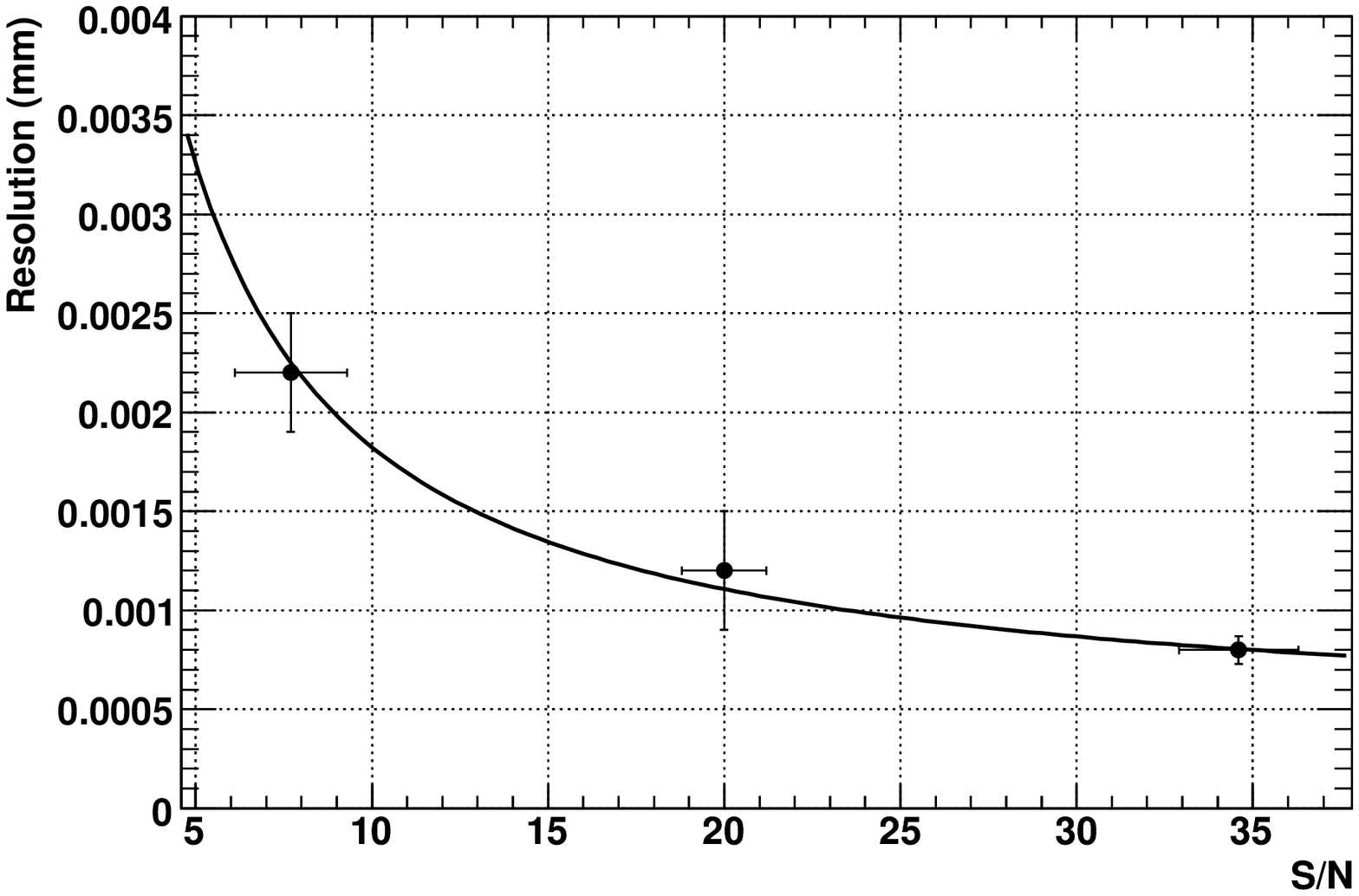,width=7.0cm}}
\end{minipage}
\end{center}
\caption[]{Test results of the LDRD-SOI-1 analog pixels with a 1060~nm focused laser.
Left: cluster pulse height as a function of the substrate bias, $V_d$ (from~\cite{battaglia_nima});
the data is normalized to the pulse height measured at $V_d$ = 0.8~V. Right: single point resolution
as a function of S/N ratio obtained from pixel scans performed with the laser focused to a 5~$\mu$m spot,
for $V_{d}$ = 7~V.}
\label{fig:laser}
\end{figure}
Charge collection as a function of substrate bias has been studied on the analog pixels
by means of a 1060~nm infrared laser focused to a $\simeq$20~$\mu$m spot. Figure~\ref{fig:laser}, left shows
the pulse height in a 5$\times$5 pixel matrix centered around the spot center as a function of the applied
depletion voltage, $V_d$. The signal increases proportionally to the depletion region thickness (equivalently,
proportionally to $\sqrt{V_d}$) for $V_d<$10~V, while for larger values the signal first saturates and then
decreases for $V_d\ge$15~V. This is interpreted as due to a back-gating effect, which seems to affect the
1.0~V transistor pixels at lower $V_d$ values compared to the 1.8~V pixels.

Pixel scans have been performed by focusing a laser beam to a $\simeq$5~$\mu$m Gaussian spot and by
shifting it in steps of 1~$\mu$m along the pixel rows using a stepping motor with a positioning accuracy of 0.1~$\mu$m. 
The pixel spatial resolution could thus be estimated from the spread of the cluster position reconstructed
by means of a center of gravity algorithm at each point in the scan. The measurements were repeated for different
signal-to-noise ratio (S/N) values, obtained by varying the laser intensity. Figure~\ref{fig:laser}, right
shows the results obtained for $V_d$ = 7~V, where data points and error bars represent respectively the mean
and r.m.s. of the distribution of the differences between the laser spot position and the reconstructed cluster
position. As expected, the measured resolution scales as the inverse of the S/N, and a single point resolution
of 1~$\mu$m is obtained for S/N ratios of 20 or larger.

\begin{figure}[t]
\begin{center}
\epsfig{file=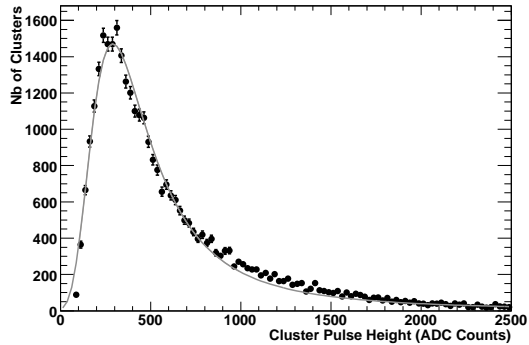,width=7.0cm}
\end{center}
\caption[]{Cluster pulse height distribution in the LDRD-SOI-1 1.8~V analog pixels obtained on the
1.35~GeV electron beam at the LBNL Advanced Light Source for a substrate bias $V_d$ = 10~V (from~\cite{battaglia_nima}).}
\label{fig:als}
\end{figure}

The response of the analog pixels to high-momentum charged particles has been tested on 
the 1.35~GeV electron beam extracted from the booster of the LBNL Advanced Light Source
(ALS). Results obtained with the analog pixels have been reported in detail in~\cite{battaglia_nima}
and are here briefly summarized. Figure~\ref{fig:als} shows the cluster pulse height distribution
obtained on the 1.8~V pixels for a depletion voltage $V_d$ = 10~V. The distribution is well fitted by
a Landau function. Consistently with the measurements performed with infrared lasers, the cluster pulse height
was found to increase up to $V_d$ = 10~V, while at 15~V the cluster signal and the efficiency of the chip
decreased. The pixel multiplicity in a cluster was found to slightly decrease with increasing $V_d$, as
expected from an increased electric field in the detector substrate. A S/N ratio of up to 15 was measured
with the 1.8~V analog section for 5~V$\le V_d \le$15~V.

\subsection{Digital Pixels}
The ALS 1.35~GeV electron beam was also used to test the response of the LDRD-SOI-1 digital pixels.
The pixels are triggered directly by the 1~Hz beam extraction signal and are latched and read
out after an integration time of 10~$\mu$s. One event is thus acquired per beam spill. For each data
set, the first 100 events are used in order to flag noisy pixels, recognized as those which are active
in more than 25 events. Furthermore, control data sets have been acquired without beam in order to
account for possible background due to noisy pixels which survive the masking and the cluster
reconstruction cuts applied in the off-line data analysis.

\begin{figure}[ht!]
\begin{center}
\begin{minipage}{.49\linewidth}
\centerline{\epsfig{file=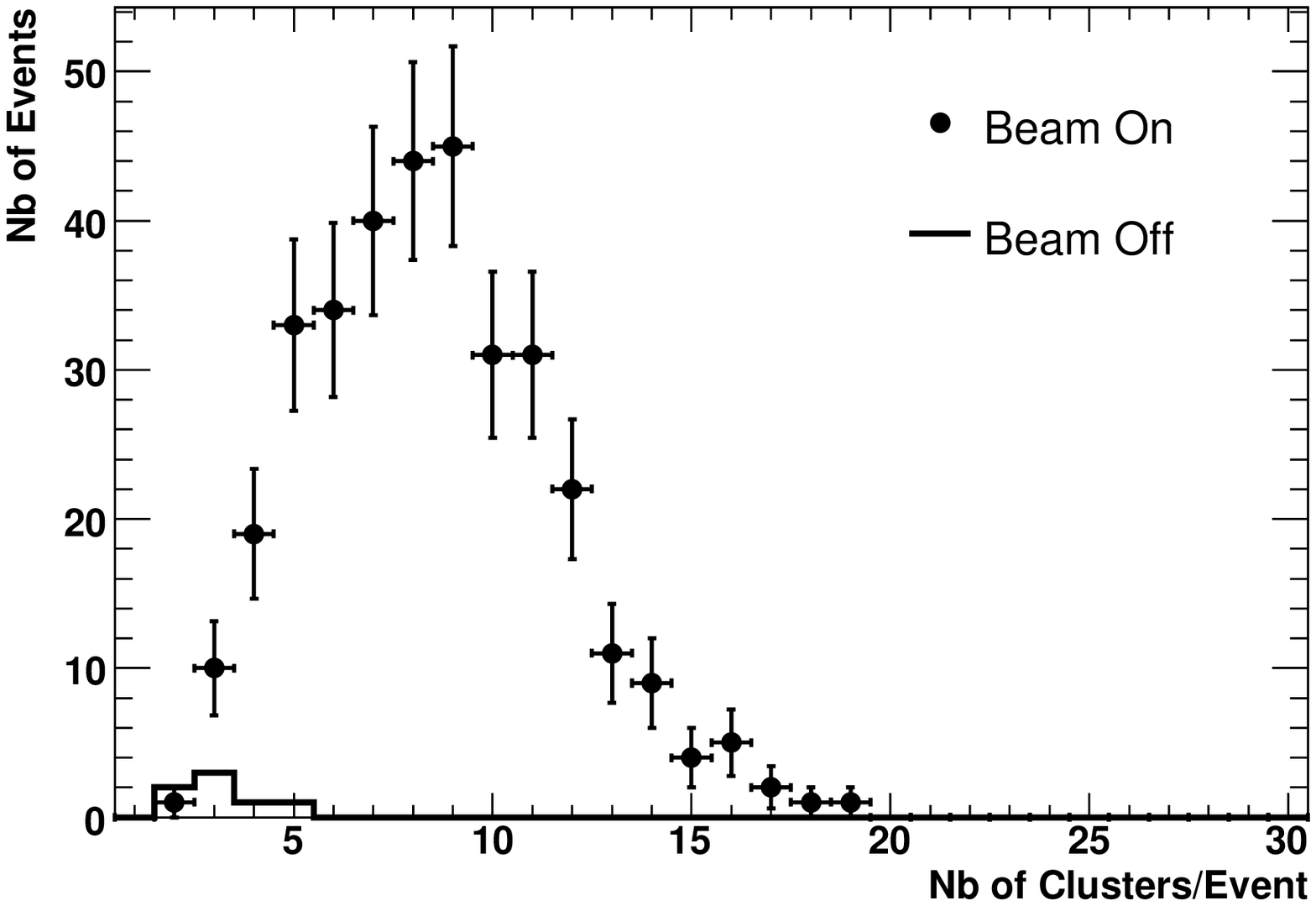,width=7.0cm}}
\end{minipage}
\hfill
\begin{minipage}{.49\linewidth}
\begin{tabular}{|c|c|c|c|}
\hline  \textbf{$V_d$} & \textbf{$\frac{Nb. Clusters}{Event}$} & 
\textbf{$\frac{Nb. Clusters}{Event}$} & \textbf{$<$Nb Pixels$>$} \\ 
\textbf{(V)}           & \textbf{beam on}                      
& \textbf{beam off}    &\textbf{in Cluster}       \\ 
\hline
20                     & 3.7$\pm$0.1         
& 0.02                 & 1.78                     \\ 
25                     & 5.3$\pm$0.1                                  
& 0.03                 & 1.32                     \\
30                     & 4.7$\pm$0.1                                  
& 0.03                 & 1.26                     \\
35                     & 4.2$\pm$0.1                                  
& 0.02                 & 1.14                     \\
\hline
\end{tabular}
\end{minipage}
\end{center}
\caption{Beam test results of the LDRD-SOI-1 digital pixels. Left: hit multiplicity distribution
for 1.35~GeV electrons for a depletion voltage $V_d$ = 30~V and high beam intensity; the continuous
line shows for comparison the distribution of fake hits reconstructed in the absence of beam. Right:
average number of clusters per event recorded on the LDRD-SOI-1 digital pixels for different values
of $V_d$, as obtained from the tests performed with a reference plane in front of the detector. Data
is reported for beam on (corrected for changes in beam intensity as described in the text) and beam
off conditions. The average multiplicity in a cluster is also reported for each $V_d$ value.}
\label{fig:als_digi}
\end{figure}

Beam signals could be observed on the digital pixels only by applying a substrate bias $V_d\ge$~20~V,
and up to $V_d$ = 35~V. As seen in the previous Section, the analog pixels do not function properly
above $V_d\simeq$~15~V. The circuitry in the digital pixels is indeed active only when clocked
and is therefore less sensitive to back-gating. Furthermore, back-gating possibly affects the
analog threshold of the in-pixel comparators, so that larger substrate voltages are required
in order to obtain signals large enough to be above threshold. Figure~\ref{fig:als_digi}, left
shows the hit multiplicity distribution observed with and without beam for $V_d$ = 30~V; an excess
of hits in the presence of beam is clearly visible. 

Due to the non-uniformity of the beam intensity across the various runs, dedicated data runs
have been taken with a single reference detector plane located 2~cm upstream 
from the LDRD-SOI-1 sensor. The reference plane used was a 50~$\mu$m-thin MIMOSA-5 CMOS pixel
sensor, developed by IPHC, Strasbourg~\cite{mimosa}. The MIMOSA-5 chip features a
1.7$\times$1.7~cm$^2$ active area with 512$\times$512 pixels on a 17~$\mu$m pitch.
The performance previously obtained on thin MIMOSA-5 sensors is presented in detail
in~\cite{Battaglia:2006tf} and~\cite{Battaglia:2008nj}. For each run, the number of electron hits
reconstructed on the reference plane allowed the particle flux to be monitored in the digital
pixel sector active area; the average number of hits per beam spill on the latter was then corrected
for the relative change of the beam intensity as determined by the reference plane. Results
obtained for different $V_d$ values are summarized in the Table in Fig.~\ref{fig:als_digi}, right.
From the particle flux reconstructed in the reference layer and its efficiency, as obtained from
simulation~\cite{Battaglia:2008nj}, the efficiency of the LDRD-SOI-1 digital pixels can
be estimated to be of the order of 0.3 to 0.5 for 20~V$\le V_d\le$35~V. Consistently with
what observed for the analog pixels, the average cluster multiplicity was found to decrease
as a function of increasing $V_d$, i.e. of the increasing electric field in
the detector substrate.

\subsection{Radiation Hardness Tests}
The effect of ionising radiation was preliminarly assessed on single test transistors, irradiated
with 30~MeV protons at the BASE Facility of the LBNL 88-inch Cyclotron~\cite{cyclotron}. The
irradiation was performed in steps up to a total fluence of 2.5$\times$10$^{12}$~p/cm$^2$,
corresponding to a total dose of $\sim$600~kRad. The characteristics of two test transistors,
one $p$-MOSFET and one $n$-MOSFET, were measured in-between irradiation steps. Both for the $n$-MOS
and $p$-MOS transistor, a significant shift in threshold voltage was observed as a function of
the increasing dose, the effect being larger for larger substrate biases. For the initial bias
of $V_d$ = 5~V the shift was of $\sim$100~mV after a fluence of about 1$\times$10$^{12}$~p/cm$^2$;
a similar shift was measured after the highest fluence for a reduced substrate bias of $V_d$ = 1~V.
The total threshold shift is much larger than the one expected at such doses from radiation damage
in the transistor thin gate oxide, and seems to be associated with the build-up of positive charge
in the detector thick buried oxide, which increases the back-gate potential on the MOS transistors. 

The effect of non-ionising radiation has been evaluated on the analog pixel matrix by exposing
the sensor to neutrons produced from 20~MeV deuteron breakup on a thin Be target at the LBNL 88-inch
Cyclotron~\cite{mcmahan}. Deuteron breakup produces neutrons on a continuum spectrum from $\sim$1~MeV
up to $\sim$14~MeV. The total neutron fluence was estimated to be 1.2$\times$10$^{13}$~n/cm$^2$
by means of activation foils placed behind the sensor. The sensor noise has been measured before and
after the irradiation as a function of the depletion voltage, $V_d$. All tests have been performed at
room temperature. A noise increase was observed after irradiation, varying from +25\% for $V_d$ = 5~V,
to +52\% for $V_d$ = 20~V. This is interpreted as due to radiation-induced increase of leakage current
in the sensor substrate. The noise of the irradiated chip was subsequently measured as a function
of the operational temperature between -5$^{\circ}$C and +20$^{\circ}$C, for $V_d$ = 10~V. The noise was observed
to decrease with decreasing temperature, consistently with what expected from
a leakage current dependence, and the pre-irradiation noise level could be recovered at temperatures
below +5$^{\circ}$C.

\section{Prototype in 0.20~$\mu$m Process}\label{soi2}
\subsection{Chip Design and Experimental Setup}

\begin{figure}[b]
\begin{center}
\epsfig{file=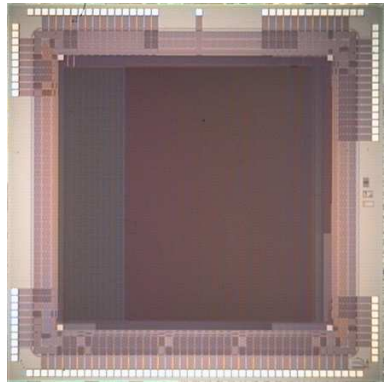,width=5cm}
\end{center}
\caption[]{Layout of the LDRD-SOI-2 chip, manufactured in the OKI 0.20~$\mu$m FD-SOI process.}
\label{fig:soi2_layout}
\end{figure}

A second prototype sensor, the LDRD-SOI-2 chip, has been designed and fabricated in 2008
in the OKI 0.20~$\mu$m FD-SOI process, optimized for low leakage current. The chip is a larger
scale prototype of 5$\times$5~mm$^2$ with an active area of 3.5$\times$3.5~mm$^2$
in which 168$\times$172 pixels of 20~$\mu$m pitch are arrayed (Fig.~\ref{fig:soi2_layout}).
The pixel matrix is subdivided into a 40$\times$172 pixel section with a simple, analog
3T architecture, and in a 128$\times$172 pixel section with a new pixel architecture providing
a binary output. In the latter, two capacitors are integrated in each pixel for in-pixel CDS,
and a digital latch is triggered by a clocked comparator with a current threshold, which is
common to the whole matrix. The chip design has been optimized in order to allow readout up
to a 50~MHz clock frequency, and the binary section is provided with multiple parallel outputs
for high frame rate. As in the LDRD-SOI-1 chip, each pixel is surrounded by a floating $p$-type
guard-ring; two floating guard-rings also separate the peripheral electronics and I/O logic
from the pixel matrix and from the pad area.

The chip is read out by a newly developed DAQ system based on a commercially available FPGA
board, deploying the Xilinx Virtex-5 device, connected to a custom-designed ADC board 
featuring five 14-bit, 100 MS/s ADCs. The chip is mounted on a proximity board which is decoupled 
from the DAQ hardware. The clock patterns from the FPGA are sent to the chip via LVDS lines,
ensuring high speed and low noise over long cable distances, while the chip analog output
is connected to one ADC input by a standard USB differential cable. The whole system
is connected to the USB bus of a PC, and controlled by a C++ based software interfaced
with {\tt ROOT} classes~\cite{root}. The data acquired by the DAQ system is sent to the
control PC via the USB bus and stored in the {\tt ROOT} format, and off-line converted into
{\tt lcio} in order to be processed through the {\tt Marlin} framework (see Section~\ref{soi1_design}). 
Details about the DAQ hardware and software can be found in~\cite{giubilato2008}.

\subsection{Preliminary Test Results of Analog Pixels}

\begin{figure}[t]
\begin{center}
\epsfig{file=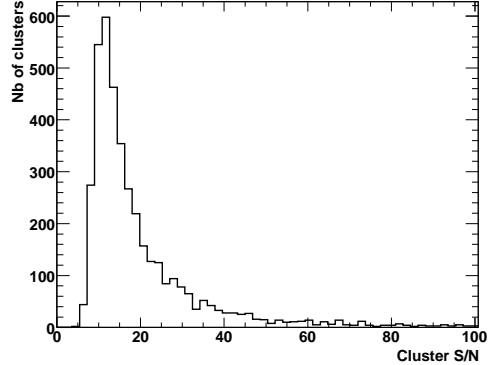,width=7.0cm}
\end{center}
\caption[]{Cluster S/N distribution obtained for the analog pixels of the LDRD-SOI-2 chip with 1.5 GeV electrons
at the LBNL ALS. The measurement was performed for a clock frequency of 12.5~MHz and a depletion voltage
$V_d$ = 2~V.}
\label{fig:soi2_als}
\end{figure}

The LDRD-SOI-2 analog pixels are being tested with a protocol similar to the one used for the LDRD-SOI-1
chip. First tests performed in the lab with infrared lasers show a different behavior of the collected
charge signal with the applied substrate bias, $V_d$. The signal pulse height increases up to $V_d$ = 2-3~V
and then rapidly saturates. The chip can be properly operated up to $V_d$ = 10~V, however, showing no significant
increase in the collected signal. This could be possibly attributed to an increased effect of back-gating due to
features of the different fabrication process or to the different guard-ring configuration, and it is
currently under investigation.

The analog pixels have been successfully tested with 1.5~GeV electrons at the LBNL ALS up to a pixel clock frequency
of 50~MHz, equivalent to an integration time of 137.6~$\mu$s for the 40$\times$172 pixel array. No significant
differences in charge collection have been observed for lower clock frequencies (i.e. longer integration times).
A S/N of $\sim$20 for reconstructed electron hits is obtained. Figure~\ref{fig:soi2_als} shows the measured cluster
S/N ratio distribution for $V_d$ = 2~V and 12.5~MHz clocking frequency. The pulse height dependence on $V_d$ is
similar to what was observed in the lab with infrared lasers. The chip leakage current and noise performance
improves with respect to the 0.15~$\mu$m process, and a single pixel noise of $\sim$30 ENC has been measured.

\section{Conclusions and Outlook}
Monolithic pixel sensors with analog and digital pixel architectures have been designed
and fabricated in SOI technologies that combine deep-submicron CMOS electronics with a
high-resistivity substrate in the same device. This is of great interest for many applications,
ranging from charged particle detection in high-energy physics to X-ray detection at
synchrotron facilities and fast beam imaging, that could benefit from the possibility
of integrating complex digital architectures with a depleted silicon substrate yielding
larger signals and ensuring a faster charge collection with respect to devices realized
in bulk CMOS processes.

A first prototype chip fabricated in a 0.15~$\mu$m process has been extensively tested using
focused infrared laser beams and a 1.35~GeV electron beam. A S/N ratio of 15 for high energy electrons
has been obtained on the analog pixels up to depletion voltages $V_d$ = 15~V, limited by back-gating
of the CMOS electronics due to the electrostatic field in the device substrate. A single point
resolution of $\simeq$~1~$\mu$m has been estimated from laser scans. The digital pixels appeared
to be less affected by back-gating and could be successfully operated at higher substrate biases,
up to $V_d$ = 35~V. Preliminary irradiation tests showed an increased effect of back-gating after
exposure to ionizing doses, due to positive charge trapping in the buried oxide. A moderate
increase of leakage current and consequently of the pixel noise was observed after exposure
to 10$^{13}$~n/cm$^2$.

A second prototype chip has been fabricated in an optimized 0.20~$\mu$m process. The chip
implements simple 3T analog pixels and an advanced binary architecture featuring in-pixel
charge storage and Correlated Double Sampling, and multiple outputs for high frame rate.
First tests of the analog pixels show an improved leakage current performance of this process.
A S/N of 20 is found for high-energy electron hits, with single pixel noise figures of 30 electrons,
for a readout frequency up to 50~MHz. Further tests of the analog section aimed at understanding
the dependence of the collected signal on the depletion voltage are on-going.

\acknowledgments
This work was supported by the Director, Office of Science, of the U.S. Department
of Energy under Contract No. DE-AC02-05CH11231. We thank the staff of the LBNL Advanced 
Light Source and 88-inch Cyclotron for their assistance and for the excellent performance 
of the machines. We are also grateful to Prof. Yasuo Arai of KEK for an effective collaboration
in the development of SOI pixel sensors.

\end{document}